\documentclass[12pt,a4paper]{article}
\pdfoutput=1
\usepackage{color}
\usepackage{amssymb,amsmath,bm,bbold}
\usepackage{epsf}
\usepackage{epsfig}
\usepackage{relsize}
\usepackage[dvipsnames]{xcolor}
\usepackage[linktoc=page,bookmarks=false,colorlinks=false,linkbordercolor=RoyalBlue,citebordercolor=ForestGreen,urlbordercolor=CornflowerBlue]{hyperref}
\usepackage{latexsym,mathrsfs,dsfont}
\usepackage[normalem]{ulem} 
\usepackage[compress]{cite}
\usepackage{graphicx}
\usepackage{url}
\usepackage{booktabs}
\usepackage{float}
\usepackage{multirow}
\usepackage{changepage}
\usepackage[hypcap]{caption, subcaption}

\usepackage[titles]{tocloft}

\usepackage{tabularx,colortbl}

\usepackage{fancyhdr}
\advance \headheight by 3.0truept       

\allowdisplaybreaks[1]

\definecolor{red}{cmyk}{0,1,1,0.4}
\definecolor{darkgreen}{rgb}{0.0,0.6,0.0}
\definecolor{cDarkGrey}{RGB}{91,91,91}
\definecolor{cGrey}{RGB}{245,243,238}
\definecolor{cBlue}{RGB}{0,110,191}
\definecolor{cLightBlue}{RGB}{214,237,252}
\definecolor{cRed}{RGB}{196,0,100}
\definecolor{cLightRed}{RGB}{254,222,237}
\definecolor{cGreen}{RGB}{0,166,80}
\definecolor{cLightGreen}{RGB}{254,222,237}
\definecolor{cOrange}{RGB}{221,74,44}
\definecolor{cLightOrange}{RGB}{255,215,210}
\definecolor{cPurple}{RGB}{93,35,125}
\definecolor{cLightPurple}{RGB}{241,230,252}
\definecolor{cYellow}{RGB}{252,191,10}
\definecolor{cISSRBlue}{RGB}{0,111,174}
\definecolor{cISSRGrey}{RGB}{167,169,172}

\newcommand{\beq}{\begin{equation}}
\newcommand{\eeq}{\end{equation}}
\newcommand{\be}{\begin{equation}}
\newcommand{\ee}{\end{equation}}
\newcommand{\bi}{\begin{itemize}}
\newcommand{\ei}{\end{itemize}}
\newcommand{\ba}{\begin{array}}
\newcommand{\ea}{\end{array}}
\newcommand{\beqa}{\begin{eqnarray}}
\newcommand{\eeqa}{\end{eqnarray}}
\newcommand{\bea}{\begin{eqnarray}}
\newcommand{\eea}{\end{eqnarray}}
\newcommand{\beqn}{\begin{eqnarray}}
\newcommand{\eeqn}{\end{eqnarray}}

\newcounter{TODO}

\newcommand{\ord}{{\cal O}}

\newcommand{\GeV}{\,\text{GeV}}

\newcommand{\vcb}{|V_{cb}|}

\newcommand{\vub}{|V_{ub}|}
\newcommand{\vts}{|V_{ts}|}

\newcommand{\epe}{\varepsilon'/\varepsilon}

\def\kpn{K^+\rightarrow\pi^+\nu\bar\nu}
\def\klpn{K_{L}\rightarrow\pi^0\nu\bar\nu}

\begin{document}

\begin{flushleft}
\end{flushleft}

\vspace{-14mm}
\begin{flushright}
  AJB-23-3
\end{flushright}

\medskip

\begin{center}
{\large\bf\boldmath
  $Z^\prime$-Tandem Mechanism for the Suppression of New Physics  in Quark Mixing   with Implications for    K, D and B Decays  
}
\\[1.0cm]
{\bf
    Andrzej~J.~Buras
}\\[0.3cm]

{\small
TUM Institute for Advanced Study,
    Lichtenbergstr. 2a, D-85748 Garching, Germany \\[0.2cm]
Physik Department, TU M\"unchen, James-Franck-Stra{\ss}e, D-85748 Garching, Germany
}
\end{center}

\vskip 0.5cm

\begin{abstract}
  \noindent
  $Z^\prime$ models belong to the ones that can most easily explain the
    anomalies in $b\to s \mu^+\mu^-$ transitions.
    However, such an explanation   by a single $Z^\prime$ gauge boson, as done in the literature, is severly   constrained by the $B^0_s-\bar B_s^0$ mixing.  Also the recent finding,
    that the mass differences 
  $\Delta M_s$, $\Delta M_d$, the CP-violating parameter $\varepsilon_K$, and the mixing induced
  CP-asymmetries $S_{\psi K_S}$ and  $S_{\psi \phi}$ can be simultaneously
  well described within the SM without new physics (NP) contributions, is  a challenge
  for $Z^\prime$ models with a single $Z^\prime$ contributing at tree-level to quark mixing. We point out
  that including a second $Z^\prime$ in the model allows to eliminate simultaneously tree-level   contributions to the five $\Delta F=2$  observables used in the determination of the CKM parameters while leaving the room for NP
  in $\Delta M_K$ and $\Delta M_D$. The latter one can be removed at the price of infecting $\Delta M_s$ or $\Delta M_d$ by NP which is presently  disfavoured.
  This pattern is transparently seen using the new
 mixing matrix for $Z^\prime$ interactions with quarks.
  This strategy allows   significant tree-level contributions to $K$, $B_s$
    and $B_d$ decays thereby allowing to explain the existing anomalies in $b\to s\mu^+\mu^-$ transitions and the anticipated anomaly in the ratio $\epe$ much easier than in $Z^\prime$-Single scenarios.
  The proposed
  $Z^\prime$-Tandem mechanism bears some similarities to the GIM mechanism
  for the suppression of the FCNCs in the SM  with the role of the
  charm quark played here by the second $Z^\prime$. However, it differs from the latter profoundly in that only NP contributions to quark mixing are eliminated at tree-level. 
  We discuss briefly the implied flavour patterns in  $K$ and $B$ decay
  observables in this NP   scenario. 
 \end{abstract}

\thispagestyle{empty}
\newpage
\setcounter{page}{1}

%
%
%
\section{Introduction}

It has recently been demonstrated in \cite{Buras:2022wpw} that the
quark mixing observables
\be\label{loop}
\varepsilon_K,\qquad \Delta M_s,\qquad \Delta M_d, \qquad S_{\psi K_S}, \qquad
  S_{\psi \phi}
\ee
can be simultaneously described within the Standard Model (SM) without
any need for new physics (NP) contributions. 
As these  observables contain by now only small hadronic uncertainties and are already  well measured, this allowed to determine precisely the CKM matrix on
the basis of these observables alone \cite{Buras:2022wpw,Buras:2021nns} without the need to face the tensions in
$\vcb$ and $\vub$ determinations from inclusive and exclusive tree-level decays
\cite{Bordone:2021oof,FlavourLatticeAveragingGroupFLAG:2021npn}.
This strategy, as pointed out in \cite{Buras:2022qip}, 
avoids also, under the assumption of negligible NP contributions to these observables, the impact of NP on the values of these parameters present likely in global fits. Simultaneously it   provides SM predictions for numerous rare $K$ and $B$  branching ratios that are most accurate to date. In this manner the size of the experimentally observed  deviations from SM predictions (the pulls) can be better estimated.

These findings, following dominantly from the 2+1+1 HPQCD lattice  calculations of $B_{s,d}-\bar B_{s,d}$ hadronic
matrix elements \cite{Dowdall:2019bea}, put very strong constraints on NP models attempting to explain  a number of anomalies in $B$ and $K$ decays of which
we list only four:
\begin{itemize}
  \item
The anomalies in the low $q^2$ bin in
$B^+\to K^+\mu^+\mu^-$ ($5.1\sigma$) and $B_s\to \phi\mu^+\mu^-$ ($4.8\sigma$)
found in  \cite{Buras:2022qip} by means of the strategy of \cite{Buras:2022wpw}. Both branching ratios
are suppressed relative to the SM predictions.
\item
  Possible anomalies in the ratio $\epe$ and $\Delta M_K$. Despite some controverses, it is likely that the SM prediction for $\epe$ has to be enhanced by NP
  to agree with data \cite{Buras:2022cyc}. For $\Delta M_K$ most recent SM results from  the RBC-UKQCD
  collaboration \cite{Bai:2018mdv,Wang:2022lfq}  are significantly larger than the data although
  due to large uncertainties this deviation is only around $2.0\,\sigma$.
  \end{itemize}
The question then arises which NP could explain these anomalies without
destroying good agreement of the SM with the experimental  data on the observables in (\ref{loop}).

It is well known that leptoquarks do not contribute to quark mixing observables
at tree-level and do not destroy the agreement of the SM with data in question
although it is advisable to avoid models that contain left-right operators
at the one-loop level. However, in the presence of lepton flavour universality in $b\to s\ell^+\ell^-$ transitions, as
indicated by the recent LHCb data \cite{LHCb:2022qnv,LHCb:2022zom}, leptoquarks cannot explain the observed suppression of
$B^+\to K^+\mu^+\mu^-$ and $B_s\to \phi\mu^+\mu^-$ 
branching ratios below the SM predictions
without violating the experimental  upper bound on the $K_L\to e^+\mu^-$ branching ratio. Similarly, leptoquarks cannot provide any significant enhancement
of $\epe$ without violating experimental upper bounds on a number of rare $K$ decay branching ratios \cite{Bobeth:2017ecx}.

The next possibility is a $Z^\prime$ gauge boson which
does not have these problems but it contributes to the observables in (\ref{loop}) at tree-level and has to face strong constraints from them.
However, as demonstrated recently in \cite{Aebischer:2023mbz}, by choosing the $Z^\prime$ coupling $\Delta^{sd}_L(Z^\prime)$ to be imaginary, tree-level $Z^\prime$ contributions to $\varepsilon_K$ can be eliminated. Simultaneously  the suppression of $\Delta M_K$, as suggested by the RBC-UKQCD result, a sizable  enhancement of $\epe$ and  a large impact of NP on  rare $K$ decays is possible, moreover in a correlated manner. This pattern is stable under renormalization group effects for a restricted range of the values of the coupling $\Delta^{sd}_L(Z^\prime)$ making this scenario rather predictive \cite{Aebischer:2023mbz}.

In the present paper we want to point out that the latter strategy
does not eliminate tree-level NP contributions to the remaining observables
in (\ref{loop}). The reason is that $\Delta M_s$ and $\Delta M_d$ are governed
by the absolute values of the mixing amplitudes 
\be
M^{bs}_{12}= (M^{bs}_{12})^{\rm SM}+M^{bs}_{12}(Z^\prime),\qquad M^{bd}_{12}= (M^{bd}_{12})^{\rm SM}+M^{bd}_{12}(Z^\prime)
\ee
and  not by their imaginary parts as is the case of $\varepsilon_K$. Analogous
comment applies to  $S_{\psi K_S}$ and $S_{\psi \phi}$. Therefore
in order  to remove NP contributions to observables in $B^0_{s,d}-\bar B^0_{s,d}$
systems at tree-level, we have to remove $M^{ij}_{12}(Z^\prime)$ completely at this level, while keeping the $Z^\prime bs$ and $Z^\prime bd$ couplings non-zero with the first required for the explanation of the observed anomalies in $b\to s\mu^+\mu^-$
transitions.

This is not possible with a single $Z^\prime$ and as illustrated in a recent analysis in 331 models in \cite{Buras:2023ldz}  some amount of NP in the observables in (\ref{loop}), at the level of $5\%$, has to be admitted to have a chance to address properly
the observed anomalies. This is still allowed in view of the remaining hadronic uncertainties, but  if one day the constraints from $\Delta F=2$
processes will become even tighter than they are at present, the 331 models and other
$Z^\prime$ models with a single $Z^\prime$ gauge boson  will likely
fail in this context.

In the present paper we want to propose a new mechanism for the suppression
of $Z^\prime$ contributions to  $\Delta F=2$
observables in (\ref{loop}) at the tree-level that  allows  simultaneously significant tree-level contributions to $K$, $D$ and $B$ decays, including new CP-violating effects.
We simply add a second $Z^\prime$ gauge boson which eliminates the tree level  contributions
of the original $Z^\prime$ to the observables in (\ref{loop}) while still
contributing together with the original $Z^\prime$ to $\Delta F=1$ observables.
In the proposed $Z^\prime$-Tandem framework the determination of the CKM parameters is separated from the determination of NP parameters as follows.
\begin{itemize}
\item
  CKM parameters are determined from quark mixing observables only. As
  $Z^\prime_1$ and $Z^\prime_2$ collaborate to remove their tree-level contributions to quark mixing observables in (\ref{loop}), this allows the determination
  of CKM parameters and in turn SM  predictions for $\Delta F=1$ observables without   NP infection. 
\item
  NP parameters describing $Z^\prime_{1,2}$ interactions with quarks and present
    in the hermitian and unitary mixing matrices proposed  recently
  \cite{Buras:2023fhi},  are determined exclusively from $\Delta F=1$ observables.   Simultaneously the two $Z^\prime$ gauge bosons collaborate in the explanation of the existing anomalies in $\Delta F=1$ decays. Due to a very strong suppression   of their contributions to all FCNC observables at the one-loop level  \cite{Buras:2023fhi}, the
  main arena for the $Z^\prime$-Tandem are  $\Delta F=1$ decays with their
  tree-level contributions only,  simplifying thereby the phenomenology.
\end{itemize}
This picture may appear to some flavour researchers as being too idealistic, but it could turn out one day to be a  good approximation of flavour changing
phenomena investigated with the help of rare decays. It should also be
clear that the proposed $Z^\prime$-Tandem is meant to represent the lightest
new particles of a complete UV completion of the SM that will include
surely a scalar system necessary to break spontaneously the gauge symmetries
represented by the the two massive $Z^\prime$ gauge bosons. Moreover, the
cancellation of gauge anomalies will surely require the introduction of new
vector-like havey fermions and one will have to make sure that the presence
of these new particles has only small impact on $\Delta F=2$ processes.
Whether such a UV completion can be constructed remains to be seen but I hope that the present paper will  motivate model builders to search for such a UV
completion.

Our paper is organized as follows. In Section~\ref{Tandem} we introduce the
$Z^\prime$-Tandem in question and derive  conditions on the couplings and masses of these two
$Z^\prime$ gauge bosons which allow us to achieve our goal. In this context
the 
mixing  matrix for $Z^\prime$ interactions with quarks, presented
recently by us \cite{Buras:2023fhi}, turns out to be crucial.
In Section~\ref{Examples} the impact of this tandem  on selected 
rare $K$ and $B$ meson decays is discussed leaving a detailed numerical analysis
for the future.
We conclude in Section~\ref{LAST}.

\boldmath
\section{$Z^\prime$-Tandem}\label{Tandem}
\unboldmath
  We add then a second $Z^\prime$ with
  appropriate quark couplings and its mass so that it cancels the contributions
  of the first $Z^\prime$ to  $\Delta F=2$ observables
 at tree-level. While at first sight one would think that several
  additional free parameters are added to this system in this manner, lowering its predictive
  power, in fact as far as quark flavour violating couplings are concerned there are none, because of the requirement
  of the removal of NP contributions to quark mixing at tree-level. However, new flavour conserving quark couplings of the new gauge boson and its  lepton couplings enter the game implying thereby a rich phenomenology. In particular
  they help in the explanation of various anomalies.

  Indeed, let us denote the quark couplings of these two gauge bosons by
  \be
\Delta_L^{ij}(Z_1^\prime)=|\Delta_L^{ij}(Z_1^\prime)|e^{i\phi^{ij}_1}, \qquad
\Delta_L^{ij}(Z_2^\prime)=|\Delta_L^{ij}(Z_2^\prime)|e^{i\phi^{ij}_2}\,,
\ee
where $(i,j)$ are quark flavour indices, either for the down-quarks or the up-quarks.
Then the two conditions for the removal of the $Z^\prime_{1,2}$ contributions to
$M^{ij}_{12}$ at tree-level read as follows
  \be\label{c1}
 \boxed{ \frac{|\Delta_L^{ij}(Z_1^\prime)|}{M_1}=\frac{|\Delta_L^{ij}(Z_2^\prime)|}{M_2}\,,
  \qquad
  \phi^{ij}_2=\phi^{ij}_1+90^\circ\,,\qquad i\not=j}
  \ee
 with $M_{1,2}$ being the masses of $Z^\prime_{1,2}$.
As  both contributions to $M^{ij}_{12}$   are given at tree-level by
\be 
M^{ij}_{12}(Z^\prime_1)=\left[\frac{\Delta_L^{ij}(Z_1^\prime)}{M_1}\right]^2\,,\qquad
M^{ij}_{12}(Z^\prime_2)=\left[\frac{\Delta_L^{ij}(Z_2^\prime)}{M_2}\right]^2\,,
\ee
it is evident that the difference between the phases $\phi^{ij}_1$ and $\phi^{ij}_2$ by $90^\circ$ assures the cancellation of these two contributions to $M^{ij}_{12}$.

However, as demonstrated in \cite{Buras:2023fhi}, each of these matrices
depends on only two mixing angles and two phases and it is not possible with
two $Z^\prime$ gauge bosons to remove NP from all $\Delta F=2$ observables, but
fortunately this can be done for the ones in (\ref{loop}).

In order to demonstrate it let us write down the explicit expressions for the matrices
$\hat\Delta(Z^\prime_1)$ and $\hat\Delta(Z^\prime_2)$  \cite{Buras:2023fhi}
\be\label{B1}
\hat\Delta (Z^\prime_1)= \begin{pmatrix}
             1-2 s_1^2s_2^2  & -2 s_1^2 s_2 c_2 e^{-i(\delta_1-\delta_2)}  & -2s_1 s_2 c_1  e^{-i\delta_1} \\ - 2 s_1^2 s_2 c_2 e^{i(\delta_1-\delta_2)}
             & 1-2 s_1^2c^2_2 & -2s_1c_1c_2  e^{-i\delta_2}     \\

-2s_1 s_2 c_1  e^{i\delta_1}	     &   -2s_1c_1c_2        e^{i\delta_2}             & 1-2 c_1^2
                \end{pmatrix},
\ee

\be\label{B2}
\hat\Delta (Z^\prime_2)= \begin{pmatrix}
             1-2 \tilde s_1^2\tilde s_2^2  & -2 \tilde s_1^2 \tilde s_2 \tilde c_2 e^{-i(\phi_1-\phi_2)}  & -2\tilde s_1 \tilde s_2 \tilde c_1  e^{-i\phi_1} \\ - 2 \tilde s_1^2 \tilde s_2 \tilde c_2 e^{i(\phi_1-\phi_2)}
             & 1-2 \tilde s_1^2\tilde c^2_2 & -2\tilde s_1\tilde c_1\tilde c_2  e^{-i\phi_2}     \\

-2\tilde s_1 \tilde s_2 \tilde c_1  e^{i\phi_1}	     &   -2\tilde s_1\tilde c_1\tilde c_2        e^{i\phi_2}             & 1-2 \tilde c_1^2
                \end{pmatrix}\,,
\ee
with $s_i,\,c_i,\,\tilde s_i,\, \tilde c_i$ standing for sines and cosines of the
mixing angles.

It is evident from these matrices that once the $bs$ and $bd$ couplings are
determined, the coupling $sd$ is also determined as discussed in
\cite{Buras:2023fhi}, a property known also from 331 models \cite{Buras:2012dp}.
One finds then that once the second relation in (\ref{c1}) is used for the phases $bs$ and $bd$, there is  no difference between the $sd$ phases 
of $Z^\prime_1$ and  $Z^\prime_2$ gauge bosons so that in this case there is no cancellation
of their contributions to $\Delta S=2$ observables like $\varepsilon_K$.  Fortunately, in this
case one can use the idea of  \cite{Aebischer:2023mbz} and set
\be\label{ABK}
\boxed{\delta_2-\delta_1=90^\circ, \qquad \phi_2-\phi_1=90^\circ.}
\ee
This means that $Z^\prime_1$ and  $Z^\prime_2$ do not collaborate in this case to remove their contributions to $\varepsilon_K$ because they can do it separately by themselves.
However, they collaborate to remove their contributions to $\Delta M_d$
and $\Delta M_s$ through the relations
\be\label{c4}
\boxed{\phi_1=\delta_1+90^\circ,\qquad \phi_2=\delta_2+90^\circ}
\ee
and
\be\label{c5}
 \boxed{ \frac{|\Delta_L^{bd}(Z_1^\prime)|}{M_1}=\frac{|\Delta_L^{bd}(Z_2^\prime)|}{M_2}\,,
\qquad
\frac{|\Delta_L^{bs}(Z_1^\prime)|}{M_1}=\frac{|\Delta_L^{bs}(Z_2^\prime)|}{M_2}\,.}
\ee

The relations (\ref{ABK}, (\ref{c4}) and (\ref{c5}) imply that we have the following free NP  parameters in the quark sector to our disposal
\be
\boxed{\delta_1, \qquad s_1, \qquad s_2, \qquad  M_1,\qquad  M_2\,}
\ee
and the remaining phases given in terms of $\delta_1$ as follows
\be\label{Phases}
\delta_2=\delta_1+90^\circ,\qquad \phi_1= \delta_1+ 90^\circ,\qquad \phi_2=\delta_1+180^\circ\,.
\ee
Thus in this favourite scenario we have to our disposal only one independent new complex phase which affects both $B_d$ and $B_s$ decays but not $K$ decays for
which $\delta_2-\delta_1=90^\circ$.

The following comments should be  made.
\begin{itemize}
\item
  The choice of cancellations made above is not the only option but presently the optimal one. It removes tree-level $Z^\prime_{1,2}$ contributions to the
  observables in (\ref{loop}). Moreover, as shown in \cite{Aebischer:2023mbz}, it provides naturally   suppression of $\Delta M_K$ as soon as the branching
  ratios for rare decays like $\kpn$ and $\klpn$ are modified by $Z^\prime_{1,2}$
  contributions. Also $\epe$ can be significantly enhanced. There are also NP contributions to $\Delta M_D$ but in view
  of hadronic uncertainties we do not expect them to be problematic.
\item
  Most importantly, the $b\to s\mu^+\mu^-$ anomalies can be explained as now
  the constraint from $\Delta M_s$ can be avoided.
\item
  The other two option in which the cancellations in question are required for
  $sd$ and $sb$ or $sd$ and $bd$ couplings will not eliminate NP contributions
  to $\Delta M_s$ or $\Delta M_d$, respectively and consequently not allowing the determination   of the CKM parameters without NP infection. But they should
  be kept in mind.
\item
  The cancellations in question implies the presence of new CP-violating phases
  which will be visible in $B$, $K$ and $D$ decays. 
\item
  This method of removing NP contributions could also be used for right-handed couplings.  But as at the one-loop level box diagrams with both bosons can be present,   one should avoid models in which  $Z_1^\prime$ has left-handed couplings
  and $Z_2^\prime$  right-handed ones or vice versa. This would generate left-right   operators whose large hadronic matrix elements and RG evolution could
  make the $Z_{1,2}^\prime$ contributions to $M^{ij}_{12}$ at one-loop level non-negligible. On the other hand the natural suppression mechanism of one-loop $Z^\prime$ contributions to all FCNC processes in \cite{Buras:2023fhi}, amounting to
  $\ord(m_b^2/M_{Z^\prime}^2)$, would likely remove these problems.
\item In the case of some signs of NP contributions to $\Delta F=2$ observables the relations  in (\ref{loop}) could be relaxed. 
\end{itemize}

It should be mentioned that in a recent UTfitter SM analysis
\cite{UTfit:2022hsi} some difficulties
in the explanation of the experimental value of $\varepsilon_K$ have been
found so that our goal to remove NP contributions to $\varepsilon_K$ could
appear to be unjustified. Yet, this can be traced back to the use by these authors of the average of $2+1$ and $2+1+1$
hadronic matrix elements in $B^0_{s,d}-\bar B^0_{s,d}$ mixings, that already
in \cite{Buras:2021nns} has been found to imply inconsistencies between observables in (\ref{loop}) within the SM. As demonstrated  in \cite{Buras:2022wpw},
these inconsistencies are removed when using 2+1+1 data from the HPQCD collaboration \cite{Dowdall:2019bea}. In my view charm contributions must be included
in the lattice calculations because at $4\GeV$, used in these calculations,  charm is a dynamical degree
of freedom and the Wilson coefficients multiplying these matrix elements
include its contributions. Another reason for the difference between
the CKM values from UTfitters and ours is the inclusion of the tree-level
values of $\vub$ and $\vcb$ by them which we do not do because of the tensions
mentioned above. More arguments for this strategy are given in \cite{Buras:2022qip},

In what follows we will look at specific decays to indicate how
the usual formulae for them are modified relative to the case of $Z^\prime$-Single
scenarios. In the final expressions  we will take the two conditions in (\ref{c1})  into account.

\section{The Impact on Rare Kaon and B Decays}\label{Examples}
\boldmath
\subsection{$\kpn$ and $\klpn$}
\unboldmath

Here we illustrate what happens in the case of  $\kpn$ and $\klpn$ decays.
Their branching ratios in the scenario in question
generalize the SM ones \cite{Buchalla:1998ba} simply as follows:
   \be \label{eq:BRSMKp}
  {\mathcal{B}(K^+\to \pi^+ \nu\bar\nu) = \kappa_+ \left [ \left ( \frac{{\rm Im} X_{\rm eff} }{\lambda^5}
  \right )^2 + \left ( \frac{{\rm Re} X_{\rm eff} }{\lambda^5}
  + \frac{{\rm Re}\lambda_c}{\lambda} P_c(X)  \right )^2 \right ] \, ,}
  \ee
  \be
  \label{eq:BRSMKL}
  {\mathcal{B}( K_L \to \pi^0 \nu\bar\nu) = \kappa_L \left ( \frac{{\rm Im}
    X_{\rm eff} }{\lambda^5} \right )^2 \, ,}
\ee
with $\kappa_{+,L}$  given by  \cite{Mescia:2007kn}
\begin{equation}\label{kapp}
\kappa_+={ (5.173\pm 0.025 )\cdot 10^{-11}\left[\frac{\lambda}{0.225}\right]^8}, \qquad \kappa_L=
(2.231\pm 0.013)\cdot 10^{-10}\left[\frac{\lambda}{0.225}\right]^8\,.
\end{equation}

In our model ($k=1,2$)
\be\label{Xeff}
X_{\rm eff} = V_{ts}^* V_{td} X_{\rm SM} + X(Z_1^\prime)+X(Z^\prime_2), \qquad
 X(Z_k^\prime)=\frac{\Delta_L^{\nu\bar\nu}(Z^\prime_k)}{g^2_{\rm SM}M_k^2}
 \Delta_L^{sd}(Z_k^\prime)
       \ee
where
\be\label{PCNNLO}
X_{\rm SM}=1.462\pm 0.017, \qquad P_c(X)=(0.405\pm 0.024)\left[\frac{0.225}{\lambda}\right]^4\,,
\ee
\be\label{gsm}
g_{\text{SM}}^2=4\frac{G_F}{\sqrt 2}\frac{\alpha}{2\pi\sin^2\theta_W}=4 \frac{G_F^2 M_W^2}{2 \pi^2} =
1.78137\times 10^{-7} \GeV^{-2}\,.
\ee

Imposing the relations in (\ref{ABK}) we obtain the scenario that is
similar to the one in \cite{Aebischer:2023mbz}.
\begin{itemize}
\item
  However, now two $Z^\prime$ gauge bosons contribute instead of one and the relevant mixing  parameters are now   correlated with the ones of $B^0_s-\bar B^0_s$ and $B^0_d-\bar B^0_d$   mixings so that correlations between anomalies in $b\to s\mu^+\mu^-$   and the ones in $K$ decays exist.
\item
  But similar to the analysis in \cite{Aebischer:2023mbz} only
  the  imaginary part of $X_{\rm eff}$ is modified and the correlation between $\kpn$ and $\klpn$ takes place on the MB branch \cite{Blanke:2009pq}, parallel to the Grossman-Nir bound \cite{Grossman:1997sk}.
  \end{itemize}

But what if one day NA62 and KOTO will find the correlation between $\kpn$ and $\klpn$ branching ratios outside the MB branch
but no sign of NP in $\varepsilon_K$ and $\Delta M_K$ will be seen? In this case within
the $Z^\prime$-Tandem scenario this would imply the other options mentioned
above in which the two gauge bosons collaborate to remove NP from
$\varepsilon_K$ and $\Delta M_K$. Let us look at this possibility as it could be realized
in the future. Imposing the relations (\ref{c1})  we find 
\be
{\rm Re} X^{\rm NP}_{\rm eff}=\frac{|\Delta_L^{sd}(Z^\prime_1)|}{g^2_{\rm SM}M_1^2}\left[
  \Delta_L^{\nu\bar\nu}(Z_1^\prime)\cos \phi_1^{sd}-\frac{M_1}{M_2} \Delta_L^{\nu\bar\nu}(Z_2^\prime)\sin \phi_1^{sd}\right]\,,
  \ee
\be
{\rm Im} X^{\rm NP}_{\rm eff}=\frac{|\Delta_L^{sd}(Z^\prime_1)|}{g^2_{\rm SM}M_1^2}\left[
  \Delta_L^{\nu\bar\nu}(Z_1^\prime)\sin\phi_1^{sd}+\frac{M_1}{M_2} \Delta_L^{\nu\bar\nu}(Z_2^\prime)\cos \phi_1^{sd}\right]\,.
\ee

Let us consider the following cases:
\begin{itemize}
\item
  For $\phi_1^{sd}=90^\circ$ only $Z^\prime_1$ contributes to the imaginary part of  $X_{\rm eff}$ but now not only NP contribution to $\varepsilon_K$ is eliminated
  but also the one to $\Delta M_K$. This could turn out to be necessary if
  the RBC-UKQCD calculations will be modified and the agreement with the data for $\Delta M_K$ will be obtained. Moreover, the real part of 
  $X_{\rm eff}$ is modified by the presence of $Z_2^\prime$ so that the correlation between $\kpn$ and $\klpn$ branching ratios takes place now {\em outside} the MB branch. It can take place on both sides of this branch dependently on the sign
  of $\Delta_L^{\nu\bar\nu}(Z_2^\prime)$. For the positive (negative) sign it is
  below (above) this branch. One can find it easily by inspecting the formulae
  above taking into account that both $\lambda_c$ and $V_{ts}$ have negative
  values.
  \item
  For $\phi_1^{sd}=0$, $Z_1^\prime$ contributes only to ${\rm Re}X_{\rm eff}$, while
  $Z_2^\prime$ contributes only to ${\rm Im}X_{\rm eff}$. Again the correlation
  between the two branching ratios is outside the MB branch. This time the sign
  of $\Delta_L^{\nu\bar\nu}(Z_1^\prime)$ matters. For the positive (negative) sign it is
  above (below) this branch.
\item
  Finally for any $\phi_1^{sd}$ different from $0^\circ$, $90^\circ$, $180^\circ$ and  $270^\circ$ and $M_1\not= M_2$ both gauge bosons
  contribute to real and imaginary parts of $X_{\rm eff}$ and again the correlation in question takes place outside the MB branch dependent on the values of neutrino couplings and the value of of $\phi_1^{sd} \in [ 0\,,2\pi]$. 
\end{itemize}

We can next go one step further and require a symmetry between the two ${\rm U(1)}$ gauge groups which could be called {\em Twins-Scenario}:
\be\label{twin}
M_1=M_2,\qquad \Delta_L^{\nu\bar\nu}(Z_1^\prime)=\Delta_L^{\nu\bar\nu}(Z_2^\prime),
\ee
which implies
\be
{\rm Re} X^{\rm NP}_{\rm eff}=\frac{|\Delta_L^{sd}(Z^\prime_1)|}{g^2_{\rm SM}M_1^2}
  \Delta_L^{\nu\bar\nu}(Z_1^\prime)\left[\cos\phi_1^{sd} -\sin \phi_1^{sd}\right]\,,
  \ee
\be
{\rm Im} X^{\rm NP}_{\rm eff}=\frac{|\Delta_L^{sd}(Z^\prime_1)|}{g^2_{\rm SM}M_1^2}
  \Delta_L^{\nu\bar\nu}(Z_1^\prime)\left[\sin \phi_1^{sd}+\cos \phi_1^{sd}\right]\,.
\ee
In this particular case there are no new free parameters relative to the
 $Z^\prime$-Single scenario.

Let us summarize. Presently, the first favourite  choice in (\ref{ABK}) implies the correlation of the two branching ratios on the MB branch. If this will not
turn out to be the case, both because of future NA62 and KOTO results and the
agreement of the $\Delta M_K$ in the SM with the data, other options will have to be considered.  In particular  
the cancellation of NP contributions to $K^0-\bar K^0$
mixing with the help of two neutral gauge bosons would have to be  then invoked.
This would allow to obtain 
 the correlations between $\kpn$
and $\klpn$ branching ratios {\em outside} the MB branch. Only for very special
values of the masses and couplings and the phase $\phi_1$ this will not be the case. One example is the {\em Twins-Scenario} in (\ref{twin}) with  $\phi_1^{sd}=45^\circ$.
In this case ${\rm Re} X^{\rm NP}_{\rm eff}$ vanishes.
But this is a very special case and 
finding one
day the experimental values of these two branching ratios outside
the MB branch, while no NP effects in $\varepsilon_K$ and $\Delta M_K$, could be a hint for two $Z^\prime$ gauge bosons at work and not only one. While such correlations can also take
place in the presence of both left-handed and right-handed couplings \cite{Blanke:2009pq}, scenarios of that type could generate sizable NP contributions to $\Delta F=2$ observables at one-loop level which we want to avoid.
Correlations with other decays, both $K$ and  $B$ decays, would help in this respect, in particular in the context of specific UV completions that include
the $Z^\prime$-Tandem in question.

\boldmath
\subsection{$b\to s\mu^+\mu^-$ Transitions}
\unboldmath
As demonstrated above, in this case both $Z_1^\prime$ and $Z_2^\prime$ are required
to remove their tree-level contributions to $B^0_{s,d}-\bar B^0_{s,d}$ mixings. The
usual formulae for the Wilson coefficients in $Z^\prime$-Single models \cite{Buras:2012jb}, that enter the discussion of the $b\to s\mu^+\mu^-$ anomalies,  are now
modified.
NP contributions to the Wilson coefficients $C_9$ and $C_{10}$ are given now by
\begin{align}
 a\, C^{\rm NP}_9 &=
 \frac{\Delta_L^{sb}(Z_1^\prime)\Delta_V^{\mu\bar\mu}(Z_1^\prime)}{M^2_1}+
 \frac{\Delta_L^{sb}(Z_2^\prime)\Delta_V^{\mu\bar\mu}(Z_2^\prime)}{M^2_2}\,,
 \\
a\, C^{\rm NP}_{10} &=
 \frac{\Delta_L^{sb}(Z_1^\prime)\Delta_A^{\mu\bar\mu}(Z_1^\prime)}{M_1^2}+
 \frac{\Delta_L^{sb}(Z_2^\prime)\Delta_A^{\mu\bar\mu}(Z_2^\prime)}{M_2^2}
   \end{align}
\noindent
with 
\be
a=\frac{1.725\cdot 10^{-9}}{\rm GeV^2}\frac{\vts}{41.9\cdot 10^{-3}}e^{i\beta_s}, \qquad \beta_s=-1^\circ.
\ee
For $C^\prime_9$ and $C^\prime_{10}$ one has to replace
$\Delta_L^{sb}(Z^\prime_k)$ by  $\Delta_R^{sb}(Z^\prime_k)$.

Imposing the relations (\ref{c1})  we find
\be
 a\,C^{\rm NP}_{9} =
 \frac{|\Delta_L^{sb}(Z_1^\prime)|}{M_1^2}e^{i\phi_1^{sb}} \left[\Delta_V^{\mu\bar\mu}(Z_1^\prime)+i\frac{M_1}{M_2}\Delta_V^{\mu\bar\mu}(Z_2^\prime)\right]\,,
   \ee
   \be
a\, C^{\rm NP}_{10} =
\frac{|\Delta_L^{sb}(Z_1^\prime)|}{M_1^2}e^{i\phi_1^{sb}} \left[\Delta_A^{\mu\bar\mu}(Z_1^\prime)+i\frac{M_1}{M_2}\Delta_A^{\mu\bar\mu}(Z_2^\prime)\right].
\ee

It should be emphasized that the $Z^\prime$-Tandem not only allows to eliminate
or reduce significantly the $\Delta F=2$ constraints but can easier explain
the anomalies in $B^+\to K^+\mu^+\mu^-$  and $B_s\to \phi\mu^+\mu^-$ than
it is possible in $Z^\prime$-Single scenarios like the 331 models
considered recently  in \cite{Buras:2023ldz}. This is in  particular the case when  the two $Z^\prime$ contributions to $C^{\rm NP}_{9}$ collaborate to obtain the
 experimental value of this Wilson coefficient. In this context
lepton couplings of both gauge bosons play an important role. They 
allow to arrange  easier the measured  ratio
of $C^{\rm NP}_{9}$ and $C^{\rm NP}_{10}$  than it is possible in $Z^\prime$-Single
scenarios.

It is also evident that generally  there will be new
CP-violating effects in $b\to s \mu^+\mu^-$ transitions that can be tested through seven angular  asymmetries $A_3,A_4,A_5,A_6^s, A_7, A_8, A_9$, in  $B\to K^*\mu^+\mu^-$ and $B_s\to\phi\mu^+\mu^-$ decays \cite{Bobeth:2008ij,Altmannshofer:2008dz} which allow the distinction between various models as stressed in \cite{Alok:2017jgr}. In the case of
$B\to K\mu^+\mu^-$ there is only one such asymmetry.
Only for very special
values of the masses and couplings and the phase $\phi_1$ the Wilson coefficients $C_9$ and $C_{10}$ will remain real as in the SM.  One example is the {\em Twins-Scenario} in (\ref{twin}) extended to muon couplings for which the formulae above simplify as follows
\be
 a\,C^{\rm NP}_{9} =
 \frac{|\Delta_L^{sb}(Z_1^\prime)|}{M_1^2}e^{i\phi_1^{sb}} \Delta_V^{\mu\bar\mu}(Z_1^\prime)\left[1+i\right]\,,
   \ee
\be
 a\,C^{\rm NP}_{10} =
 \frac{|\Delta_L^{sb}(Z_1^\prime)|}{M_1^2}e^{i\phi_1^{sb}} \Delta_A^{\mu\bar\mu}(Z_1^\prime)\left[1+i\right]\,.
   \ee
For $\phi_1^{sb}=135^\circ$ both $C^{\rm NP}_{9}$ and $C^{\rm NP}_{10}$ are real
except for $\beta_s$ which has nothing to do with NP but with the standard defintion of the coefficients in question. Note that for this phase ${\rm Im} X^{\rm NP}_{\rm eff}$ vanishes in this scenario and there is no NP contribution to $\klpn$.

It should also be emphasized that in $Z^\prime$ models not only $C_9$ and $C_{10}$
Wilson coefficients but also their right-handed counterparts $C^\prime_9$ and $C^\prime_{10}$ could be relevant. However, to avoid left-right operators contributing
to $B^0_s-\bar B^0_s$ mixing at one-loop level it is favourable to have only one of 
these two pairs, which simplifies phenomenological analyses.

\section{Summary}\label{LAST}
In this paper, we have demonstrated that it is possible to avoid NP contributions
to $\Delta F=2$ observables with only moderate  tuning of parameters by adding to the usual
$Z^\prime$ models with a single $Z^\prime$ a second $Z^\prime$ which cancels the
contributions of the first $Z^\prime$ to these observables. It should be stressed that this cancellation
takes place for any value of quark flavour couplings involved, provided they
satisfy two relations given in (\ref{c1}). In this manner various anomalies
in $K$ and $B$ decays and also in the ratio $\epe$ can be explained without any worry about
the observables in (\ref{loop}) which are already well described by the SM.
This is also supported by the strong suppression of one-loop contributions
demonstrated recently in \cite{Buras:2023fhi}. In this manner the only arena for $Z_1^\prime$  and  $Z_2^\prime$ gauge bosons are $\Delta F=1$ processes. Moreover, it is
    sufficient to include these contributions at tree-level.

This is of course an important benefit compared to the common $Z^\prime$-Single
models. Moreover, in the case of Twins-Scenario the number of free parameters is not increased. Also,  with negligible NP contributions to $\Delta F=2$ observables, CKM parameters
can be determined from the latter processes as done in  \cite{Buras:2022wpw}
so that this NP scenario has only few parameters. Indeed
one can then fix the values of the CKM parameters to \cite{Buras:2022wpw}
\be\label{CKMoutput}
{\vcb=42.6(4)\times 10^{-3}, \quad 
\gamma=64.6(16)^\circ, \quad \beta=22.2(7)^\circ, \quad \vub=3.72(11)\times 10^{-3}\,}
\ee
and use them in a global fit leaving out this time the $\Delta F=2$ observables.
With a sufficient number of observables, like the ones present in
Flavio \cite{Straub:2018kue} and HEPfit  \cite{DeBlas:2019ehy} codes, the new
parameters in (\ref{B1}) and (\ref{B2}) can be determined solely from $\Delta F=1$ processes, powerful tests of this NP scenario can be made and possible anomalies explained.
In this manner also some information on the masses $M_1$ and $M_2$ could be obtained.
Even more important would be the construction of specific UV completions
which would allow one to make more concrete predictions for FCNC observables
than in the simplified scenario presented here.

We believe that the proposal of the $Z^\prime$-Tandem scenario opens a new direction
for constructing UV completions that would facilitate the explanation of the
anomalies in $\Delta F=1$ processes without strong constraints from $\Delta F=2$
ones. Therefore, we thought that before constructing new UV completions that include the $Z^\prime$-Tandem, it was appropriate to share
these ideas with flavour community already at this stage. They
could turn out to  be useful for
studying various anomalies indicated by the experimental  data. Moreover,
this new idea can be extended to {\em S-Tandems} of two neutral scalar particles,
although in this case at the one-loop level the suppression of NP in $\Delta F=2$ observables will not be as effective as in the case of the left-handed $Z^\prime$-Tandem scenario because of enhanced matrix elements of $\Delta F=2$ scalar-scalar operators.
One could also generalize this idea to $W^\prime$, $G^\prime$ and other tandems
involving scalar and vector bosons.

  The strategy for suppressing NP contributions to $\Delta F=2$ observables proposed  here, bears some similarities to the  GIM mechanism
  \cite{Glashow:1970gm} although it
  differs from it in a profound manner. In the latter case it was
  crucial to add the fourth quark, the charm quark, in order to remove  tree-level $Z$ contributions to flavour-violating observables. Here, the role of
  of the charm quark is played by $Z_2^\prime$. However, in contrast to the GIM mechanism, which forbids all FCNC processes at tree-level within the SM, our strategy, while forbidding tree-level NP  contributions to quark mixing, allows such contributions
  to rare $B$ and $K$ decays and also to $\epe$ which seems to be required
  by the data  \cite{Buras:2022cyc}. Moreover, in contrast to the GIM
  mechanism, in this tandem scenario, at one-loop level, $Z^\prime_{1,2}$ contributions to all flavour observables in $K$ and $B_{s,d}$ systems, not only to quark mixing,  being governed by down-quark masses can be shown to
  be $\ord(m_b^2/M^2_{Z^\prime})$ and consequently negligible \cite{Buras:2023fhi}.
   Thus  tree-level contributions to $K$ and
   $B_{s,d}$ decays are the only NP contributions one has to consider simplifying significantly the phenomenology. For charm mesons, being governed  at one-loop level by   up-quark masses, in particular the top quark mass, they can be relevant but    being $\ord(m_t^2/M^2_{Z^\prime})$ they are likely small.

  What remains
  is the discovery of the $Z^\prime$-Tandem at the LHC. However, it could turn out that only the lighter $Z^\prime$ can be discovered at the LHC. Yet, if
  the $\Delta F=2$ constraints will remain as strong as they are now and various
  $b\to s \mu^+\mu^-$ branching ratios will be significantly suppressed below
  the SM predictions, within $Z^\prime$ models there does not seem to be another simple solution beyond the existence of a second $Z^\prime$ gauge boson. One
  possibility would be the inclusion of right-handed couplings in addition to left-handed ones. This allows to  suppress NP contributions from a single $Z^\prime$   gauge boson to  $\Delta F=2$ observables but requires 
  fine tuning between left-left, right-right 
  and left-right operators contributing to these observables
  \cite{Buras:2014sba,Buras:2014zga,Crivellin:2015era}. Moreover, this tuning depends on 
  hadronic matrix elements of involved operators that is avoided in the present
  strategy as only short distance contributions are involved. In the context
  of the analyses in \cite{Buras:2014sba,Buras:2014zga} this new strategy
  would allow to help, without this fine tuning,  to explain partly  the $\Delta I=1/2$ rule in case
  it would turn out to be necessary one day and to probe easier very short distance scales with the help of rare $K$ and $B_{s,d}$ decays than it is possible with $Z^\prime$-Single scenarios.
  
  The existence of the second $Z^\prime$ could also be signalled by the
  correlation of $\kpn$ and $\klpn$ branching ratios outside the MB branch
  and the necessity of NP contributions to $\epe$ but none to $\Delta M_K$
  as already mentioned earlier. In any case it will be fun to explore this
  new framework in more details in various directions in the coming years.
  
  \medskip
  
  {\bf Acknowledgements}
  The discussions with Monika Blanke, Andreas Crivellin and Peter Stangl
  are highly appreciated.
 Financial support from the Excellence Cluster ORIGINS,
funded by the Deutsche Forschungsgemeinschaft (DFG, German Research
Foundation), Excellence Strategy, EXC-2094, 390783311 is acknowledged.

\renewcommand{\refname}{R\lowercase{eferences}}

\addcontentsline{toc}{section}{References}

\bibliographystyle{JHEP}

\small

\bibliography{Bookallrefs}

\end{document}